\documentclass[twocolumn,aps,showpacs,floatfix,prc]{revtex4-1}
\usepackage[dvips]{epsfig}
\usepackage{graphicx}

\begin{document}

\title {Stability against $\alpha$ decay of some recently observed superheavy elements }

\author{Partha Roy Chowdhury}
\email{royc.partha@gmail.com} 

\author{G. Gangopadhyay}
\author{Abhijit Bhattacharyya}

\affiliation{Department of Physics, University of Calcutta, 92 A.P.C. Road, Kolkata-700 009, India}

\date{\today }

\begin{abstract}

    The probability of $\alpha$ particle emission for some recently observed superheavy nuclei (SHN) are investigated. The $\alpha$-decay half lives of SHN are calculated in a quantum tunneling model with density dependent M3Y (DDM3Y) effective nuclear interaction using theoretical and measured $Q_\alpha$ values. We determine the density distribution of $\alpha$ and daughter nuclei from the relativistic mean field theory (RMF) using FSUGold force, NL3 and TM1 parameter sets. The double folded nuclear potential is numerically calculated in a more microscopic manner using these density distributions. The estimated values of 
$\alpha$-decay half-lives are in good agreement with the recent data. We compare our results with recently detected $\alpha$-decay chains from new element with atomic number Z=117 reported by JINR, Dubna. Finally, we determine the half-lives of superheavy elements with Z=108-120 and neutron number N=152-190 to explore the long-standing predictions on the existence of an ``island of stability" due to possible spherical proton (Z$\sim 114$) and neutron (N$\sim 184$) shell closures.

\vskip 0.2cm
\noindent

\end{abstract}


\maketitle

\noindent

    The formation of superheavy nuclei (SHN) by fusion was intensively explored \cite{den00,smz01,zag01}.
The investigation of experimental data concerning fusion and fission of the SHN with Z = 108-118 together with data on survival probability of these nuclei in evaporation channels with 3-4 neutrons, revealed the fact that the hindrance due to high fission barriers causes to a
relatively higher stability of such heavy nuclear systems \cite{itk02}. The investigation of $\alpha$-decay chains of such SHN is the main tool to extract some information regarding their degree of stability and possible existence in nature. In the last decade several theoretical and experimental works \cite{gam05,hof00,mor07,oga06} were devoted to the formation of SHN and their $\alpha$-decay half-lives. The $\alpha$ decay of superheavy nuclei \cite{adndt07,poen106,gg08,gg09,deli07,zha07} is possible if the shell effect supplies the extra binding energy and increases the barrier height of fission.

      In our previous works \cite{adndt07,prc06,prc08} we showed the applicability of our calculation using DDM3Y interaction in predicting the $\alpha$ decay half lives of SHN from a direct comparison with the experimental data \cite{mor07,oga06}. In this paper, we calculate the $\alpha$-decay half-lives of new superheavy element Z=117 and its decay products recently observed in JINR, Dubna \cite{ogaprl10}. The density distribution of $\alpha$-particle and daughter nuclei are determined from the RMF theory using FSUGold force \cite{prl}, NL3 \cite{NL3} and TM1 \cite{TM1} parameter sets. The Lagrangian model NL-SV1 with the inclusion of the vector self-coupling of the $\omega$-meson was successfully used by Saldanha, Farhan and Sharma \cite{shar09} to study the properties and shell structure of the superheavy elements from Z=102-120 based on RMF+BCS calculations for an axially deformed configurations of nuclei. To include the deformation effect to some extent, we have used the $Q_\alpha$ values obtained from the force NL-SV1 \cite{shar00} for the investigation of existence of the so-called ``magic island" with extra stability. However, we have determined density distribution from RMF calculation assuming spherical nuclei. In addition, the expression of centrifugal barrier is modified to introduce an additional turning point near the origin, even for the $ l= 0$ case.


The  half lives for $\alpha$ disintegration process \cite{kps10} are calculated using WKB approximation \cite{landau} in the frame work of quantum mechanical tunneling of an $\alpha$ particle from a parent nucleus~\cite{prc06}. The details of calculation of the $\alpha$ decay half lives ($T_\alpha$) of superheavy nuclei were described in our earlier works \cite{adndt07,prc08}. The required nuclear potentials are calculated by double folding the nuclear density distribution functions of the $\alpha$ particle and the daughter nucleus with DDM3Y effective interaction. The values of parameters used in the DDM3Y interaction are kept unchanged as calculated in our previous papers \cite{prc06,npa811}. In this work, the density distributions are obtained from the nuclear wave functions ($\psi$) using the Lagrangian parameter sets FSUGold, NL3 and TMI. The microscopic $\alpha$-nucleus potential thus obtained, along with the Coulomb interaction potential and the minimum centrifugal barrier required for the spin-parity conservation, form the potential barrier. The spin-parity conservation forces a minimum angular momentum ($l$) to be carried away in the decay process. Consequently, this gives rise to the Langer modified centrifugal barrier

\begin{equation}
 V_l = \hbar^2 (l+1/2)^2 / (2\mu R^2)
\label{seqn2}
\end{equation}
\noindent
where $\mu$ is the reduced mass of the daughter and emitted nuclei system and $R$ is the distance between them. In this work, minimum centrifugal barrier is assumed by using $l=0$. The Langer modification from $l(l+1) \rightarrow (l+1/2)^2$ is a necessary transformation \cite{kelkar} while going from the one dimensional problem with $x$ ranging from $-\infty \rightarrow +\infty$ to the radial one-dimensional tunneling with $r$ ranging from $0 \rightarrow \infty$. 

In this calculations, the experimental $Q_\alpha$ are used for the estimate of $T_\alpha$ of the recently synthesized new element Z=117 in JINR, Dubna \cite{ogaprl10} and recently observed decay of $^{288,289}114$ in GSI, Darmstadt \cite{gsiprl10}. The theoretical $Q_\alpha$-values for the neutron rich isotopes of the elements with even atomic number Z=108-120 are obtained from the force NL-SV1 to search the existence of an ``island of stability" surrounded by unstable nuclear systems.

The experimental decay $Q$ values ($Q_{ex}$) have been obtained from the measured $\alpha$ particle kinetic energies $E_{\alpha}$ using the following expression 

\begin{equation}
 Q_{ex} = (\frac{A_p}{A_p-4})E_{\alpha} + (65.3 Z_p^{7/5} - 80.0 Z_p^{2/5}) \times 10^{-6} ~\rm MeV
\label{seqn3}
\end{equation}
\noindent
where the first term is the standard recoil correction and the second term is an electron shielding correction. The mass and atomic numbers of the parent nucleus are denoted as $A_p$ and $Z_p$ respectively. As the $Q_\alpha$-value appears inside the exponential integral as well as in denominator of the expression \cite{prc06} of $\alpha$ decay half lives, the entire calculation is very sensitive to Q-values. 

The RMF theory has successfully described various properties of nuclei from light to superheavy domain. The RMF calculations \cite{walecka86} are used to explore the nature of possible magic numbers using various interactions and it is shown in some calculations that the patterns of single -particle levels are significantly modified in superheavy elements \cite{bender01}. Being based on the Dirac Lagrangian density ($\mathcal{L}$), RMF is particularly suited to investigating these nuclei because it naturally incorporates the spin degrees of freedom. There are different variations of the Lagrangian density and also a number of different parameterizations in RMF namely, FSUGold force, NL3 and TM1 etc. The following Lagrangian density has recently been proposed \cite{prl} which involves self-coupling of the vector-isoscalar meson as well as coupling between the vector-isoscalar meson and the
vector-isovector meson. The corresponding parameter set is called FSUGold.

\noindent
\begin{eqnarray}
\mathcal{L}=\bar{\psi}\left (i\gamma_\mu\partial^\mu - M_N\right) \psi 
+\frac{1}{2}\left (\partial_\mu\sigma\partial^\mu\sigma - m_\sigma^2\sigma^2 \right) \nonumber \\
-\frac{1}{4}\Omega_{\mu\nu}\Omega^{\mu\nu} + \frac{1}{2}m_\omega^2\omega_\mu\omega^\mu 
- \frac{1}{4}\vec{\rho}_{\mu\nu}.\vec{\rho}^{\mu\nu} \nonumber \\
+\frac{1}{2}m_\rho^2\vec{\rho}_\mu.\vec{\rho}^\mu - \frac{1}{4}A_{\mu\nu}A^{\mu\nu}
+ g_\sigma\bar{\psi}\psi\sigma  \nonumber \\ 
 - \bar{\psi}\gamma_\mu(g_\omega \omega^\mu +\frac{g_\rho}{2}\vec{\tau}.\vec{\rho}^\mu 
+ \frac{e}{2} A^\mu\left (1 + \tau_3\right))\psi \nonumber \\
- \frac{\kappa}{3!}(g_\sigma\sigma)^3 -  \frac{\lambda}{4!}(g_\sigma\sigma)^4
+ \frac{\zeta}{4!}(g^2_\omega\omega_{\mu}\omega^{\mu})^2 \nonumber \\
+ \Lambda_v(g^2_{\rho}{\vec{\rho}}_\mu.{\vec{\rho}}^\mu)(g^2_\omega\omega_{\mu}\omega^{\mu})    
\end{eqnarray}
where,
\begin{eqnarray}
A_{\mu\nu} = \partial_\mu A_\nu - \partial_\nu A_\mu\\
\Omega_{\mu\nu} = \partial_\mu\omega_\nu - \partial_\nu\omega_\mu\\
\vec{\rho}_{\mu\nu} = \partial_\mu\vec{\rho}_\nu - \partial_\nu\vec{\rho}_\mu - 
g_\rho\left(\vec{\rho}_\mu X \vec{\rho}_\nu\right)
\end{eqnarray}
\noindent
In this work, we determine the density distributions of emitter and daughter nuclei from RMF calculations using three different parameter sets namely, FSUGold force \cite{prl}, NL3 \cite{NL3} and TM1 \cite{TM1}. 


In this short report, our primary aim is to check the stability of recently discovered \cite{ogaprl10} new superheavy element with atomic number Z=117 against $\alpha$-decay. The two new isotopes $ ^{293} 117$ and $^{294} 117$ were recently produced along with 9 more new nuclei as their decay products at JINR, Dubna \cite{ogaprl10} in fusion reaction between $^{48}$Ca and $^{249}$Bk. In this work, the $\alpha$-decay lifetimes ($\tau$) are determined within a WKB framework using density distributions for daughter and emitted nuclei from RMF calculation. In Table I the experimental ($\tau^{EXP}$) \cite{ogaprl10} and theoretical $\alpha$-decay lifetimes ($\tau^{M3Y}_{FSUGold}$, $\tau^{M3Y}_{NL3}$, $\tau^{M3Y}_{TM1}$) calculated by using density distributions from FSUGold, NL3 and TM1 parameter sets are compared for $^{293, 294}117$ and their decay products. In addition, the calculations are also done with theoretical $Q_\alpha$ values used in Ref. \cite{ogaprl10} based on macroscopic-microscopic calculations of masses of superheavy nuclei \cite{sobic10}. The upper and lower limits of calculated values are arisen due to the error encountered in measured $Q_\alpha$ values. It is clear from Table I that the three sets of calculated $\alpha$-decay lifetimes $\tau^{M3Y}_{FSUGold}$, $\tau^{M3Y}_{NL3}$, $\tau^{M3Y}_{TM1}$ using wave functions from FSUGold, NL3 and TM1 parameter sets respectively with DDM3Y interaction are close to each other. Our calculations using experimental $Q_\alpha$ are in reasonable agreement with experiments for some nuclei like $^{293}117$, $^{285}113$, $^{294}117$, $^{286}113$ etc. For the rest of the nuclei like $^{290}115$, $^{282}111$, $^{278}109$, $^{274}107$ etc. our calculations are in better agreement with theoretical lifetimes given in Ref.\cite{ogaprl10}. For example, the calculated lifetimes using FSUGold ($\sim 11.3$ seconds) is much less than measured lifetimes ($\sim 1.3$ minutes) for $^{274}107$. However, this discrepancy is removed if we employ the $Q_\alpha$ values from Ref.\cite{sobic10}. The data with higher statistics are needed to determine the lifetimes of SHN with a better accuracy. 

Recently the $\alpha$ decay chains from $^{289}114$ and $^{288}114$ are reported by Dullmann {\it et al.} \cite{gsiprl10}. In Table II the calculated $\alpha$ decay half lives ($T^{FSUGold}_{1/2}$, $T^{NL3}_{1/2}$, $T^{TM1}_{1/2}$) and experimental $\alpha$ decay half lives ($T^{EXP}_{1/2}$) are given for comparison. The half lives predicted in this work are in good agreement with the measured values for $^{289}114$ and $^{281}Ds$. For nuclei $^{285}112$ and $^{288}114$ our calculations are in reasonable agreement with experiments.

Finally, the $\alpha$-decay stability of a series of superheavy nuclei with Z=108-120 towards neutron rich domain are investigated by using $Q_\alpha$ values obtained from the force NL-SV1 \cite{shar09}. The nuclear potential are numerically determined using double folding the DDM3Y interaction with density distribution functions obtained from the FSUGold parameter set. The calculated halflives are shown in Fig. 1  with increasing neutron number N=152-190 for elements with even atomic number Z=108-114. The half life for any specific element (i.e. for fixed Z) increases with increasing neutron number. This matches the trend of higher stability towards neutron shell closure around N$\sim 184$ or 186 as predicted by many microscopic calculations. A sharp fall in $T_\alpha$ around N=184 for elements Z=108, 110, 112, 114 may indicate the signature of shell effects around N=184 or 186. A similar trend is observed in Fig. 2 for elements Z=116, 118, 120. However, as the atomic number increases the shell effect seems to be less prominent. For example, the shell effect near N=184 becomes less significant for Z=118 and 120. 

In summary,  the $\alpha$-decay halflives T$_\alpha$ of some recently observed superheavy nuclei are determined in a more microscopic approach using DDM3Y interaction and RMF wave functions obtained from three different parameter sets FSUGold, NL3 and TM1. The calculations are in reasonable agreement with the measured values of T$_\alpha$. Finally, a thorough investigations are done to search the existence of an ``island of stability" in neutron rich domain of superheavy nuclei with Z=108-120 and N=152-190 using the $Q_\alpha$ values from NL-SV1 parameter set. It seems from this calculation that the longer lived superheavy nuclei may exist due to shell effect near N$\sim 184$ with Z=110-114. However, more experimental and theoretical works are needed to be done to shed more light on this issue.\\

\begin{figure}
\includegraphics[width=3.5in]{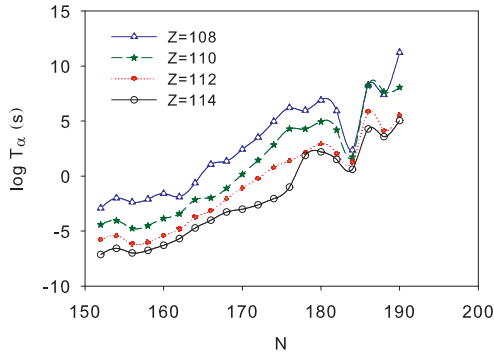}
\caption{(Color online) Half lives log T$_\alpha$ (in second) vs. neutron number (N) using density distribution from the force FSUGold and $Q_\alpha$ from NL-SV1 parameter set for Z=108-114  }
\label{fig1}
\end{figure}

\begin{figure}
\centerline{\includegraphics[width=3.5in]{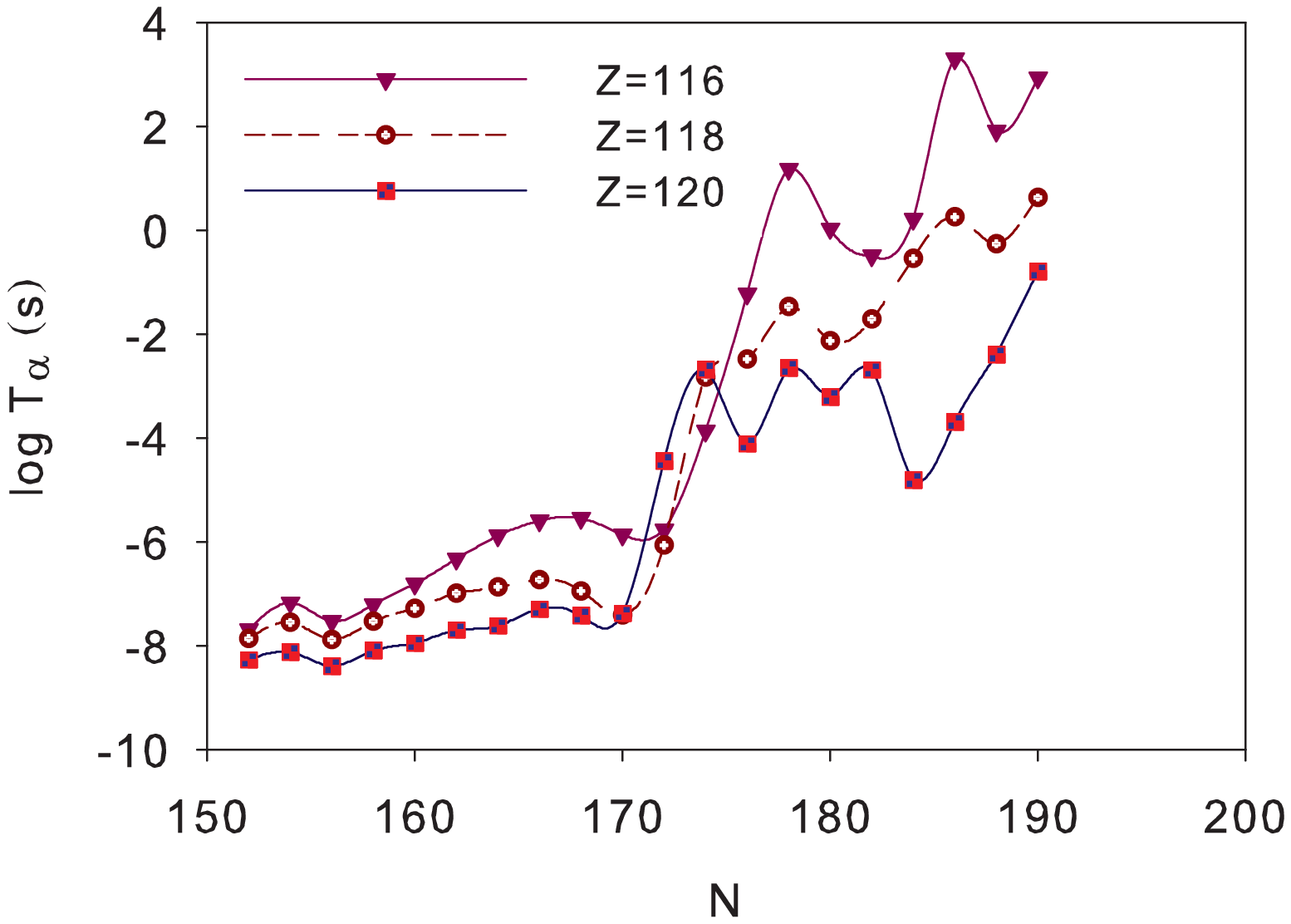}}
\caption{(Color online) Half lives log T$_\alpha$ (in second) vs. neutron number (N) using density distribution from the force FSUGold and $Q_\alpha$ from NL-SV1 parameter set for Z=116-120}
\label{fig2}
\end{figure}

\begin{table*}
\caption{Comparisons between observed ($\tau^{EXP}$) \cite{ogaprl10} and theoretical (this work) $\alpha$-decay lifetimes ($\tau^{M3Y}_{FSUGold}$, $\tau^{M3Y}_{NL3}$, $\tau^{M3Y}_{TM1}$) using measured and calculated $Q_\alpha$ given in Ref.\cite{ogaprl10}. 
The calculated $Q_\alpha$ values taken from Ref. \cite{sobic10} are given with single (*) asterisk symbol. The respective half lives given in Ref.\cite{ogaprl10} using calculated $Q_\alpha$ are also denoted by single (*) asterisk symbol.}
\begin{ruledtabular}
\begin{tabular}{cccccccccc}
Parent  & Ref. \cite{ogaprl10} & Ref. \cite{ogaprl10}  & Lifetimes  &Lifetimes&Lifetimes& Lifetimes\\
\hline 
$^AZ$   & E$_\alpha ^{expt}$ (MeV)    &$Q_\alpha~(MeV)$    &$\tau^{EXP}$ Ref.\cite{ogaprl10}    &$\tau_{FSUGold} ^{M3Y}$     &$\tau_{NL3} ^{M3Y}$      &$\tau_{TM1} ^{M3Y}$ \\
\noalign{\smallskip}\hline\noalign{\smallskip}

$^{293}117$   &	    11.03 (8)	& 11.19 (8)   &	  21~ms 	            &$	4.1^{+2.3}_{-1.5}~ms	$&$	3.5^{+2.1}_{-1.2}~ms	$&$	3.6^{+2.1}_{-1.3}~ms	$\\
$^{293}117$   & $^{*}$11.26	& 11.42          &    $^{*}$10~ms 	&$	1.2~ms	$&$	1.0~ms	$&$	1.0~ms	$\\

$^{289}115$   &	    10.31 (9)	& 10.46 (9)   &	  0.32~s 	            &$	65^{+47}_{-27}~ms	$&$	56^{+40}_{-24}~ms	$&$	57^{+41}_{-24}~ms	$\\
$^{289}115$   & $^{*}$10.48	& 10.63          &    $^{*}$0.22~s 	&$	23~ms	$&$	20~ms	$&$	20~ms	$\\

$^{285}113$   &	    9.48 (11)	& 9.62 (11)   &	  7.9~s 	            &$	3.0^{+3.4}_{-1.6}~s	$&$	2.6^{+2.8}_{-1.4}~s	$&$	2.6^{+3.2}_{-1.4}~s	$\\
$^{285}113$   &	    9.74 (8)	& 9.89 (8)     &	  7.9~s 	            &$	530^{+366}_{-213}~ms	$&$	456^{+312}_{-184}~ms	$&$	464^{+317}_{-187}~ms	$\\
$^{285}113$   & $^{*}$9.96 	& 10.11          &    $^{*}$1.2~s 	&$	0.13~s	$&$	0.11~s	$&$	0.11~s	$\\
\hline
$^{294}117$   &	    10.81 (10)	& 10.97 (10)   &	  112~ms 	            &$	52^{+46}_{-23}~ms	$&$	45^{+39}_{-20}~ms	$&$	48^{+38}_{-22}~ms	$\\
$^{294}117$   & $^{*}$11.00	& 11.16          &    $^{*}$45~ms 	&$	18~ms	$&$	15~ms	$&$	16~ms	$\\

$^{290}115$   &	    9.95 (40)	& 10.10 (10)   &	  0.023~s 	            &$	2.43^{+33.18}_{-2.24}~s	$&$	2.09^{+28.41}_{-1.93}~s	$&$	2.13^{+28.89}_{-1.96}~s	$\\
$^{290}115$   & $^{*}$10.23	& 10.38          &    $^{*}$1.0~s 	&$	0.40~s	$&$	0.34~s	$&$	0.35~s	$\\

$^{286}113$   &	    9.63 (10)	& 9.77 (10)   &	               28.3~s 	            &$	4.2^{+4.0}_{-2.0}~s	$&$	3.6^{+3.4}_{-1.8}~s	$&$	3.6^{+3.5}_{-1.7}~s	$\\
$^{286}113$   & $^{*}$9.56  	& 9.70           &    $^{*}$16~s 	&$	6.7~s	$&$	5.7~s	$&$	5.8~s	$\\

$^{282}111$   &	    9.00 (10)	& 9.13 (10)   &	  0.74~s 	            &$	70.1^{+77.3}_{-35.9}~s	$&$	59.95^{+66.01}_{-30.79}~s	$&$	60.91^{+67.05}_{-31.27}~s	$\\
$^{282}111$   & $^{*}$9.43 	& 9.57          &    $^{*}$8.1~s 	&$	3.4~s	$&$	2.9~s	$&$	3.0~s	$\\

$^{278}109$   &	    9.55 (19)	& 9.70 (19)   &	  11.0~s 	            &$	0.3^{+0.8}_{-0.2}~s	$&$	0.3^{+0.6}_{-0.2}~s		$&$	0.3^{+0.7}_{-0.2}~s		$\\
$^{278}109$   & $^{*}$9.14	            & 9.28          &    $^{*}$13~s 	&$	5.0~s	$&$	4.3~s	$&$	4.3~s	$\\

$^{274}107$   &	    8.80 (10)	& 8.94 (10)   &	  1.3~min 	            &$      11.3^{+12.1}_{-5.9}~s	$&$	9.6^{+10.4}_{-5.0}~s	$&$	9.7^{+10.5}_{-5.0}~s	$\\
$^{274}107$   & $^{*}$8.43 	& 8.56          &    $^{*}$7.4~min 	&$	3.02~min	$&$	2.57~min	$&$	2.62~min	$\\

\end{tabular}
\end{ruledtabular}
\end{table*}

\begin{table*}
\caption{Comparisons between observed \cite{gsiprl10} and theoretical (this work) $\alpha$-decay half lives ($T_{1/2}^{FSUGold}$, $T_{1/2}^{NL3}$, $T_{1/2}^{TM1}$) using measured $Q_\alpha$ \cite{gsiprl10}.}
\begin{ruledtabular}
\begin{tabular}{cccccccccc}
Parent  & Ref. \cite{gsiprl10} & Ref. \cite{gsiprl10}  & Halflives  &Halflives&Halflives& Halflives\\
\hline 
$^AZ$   & E$_\alpha ^{expt}$ (MeV)    &$Q_\alpha~(MeV)$    &$T_{1/2}^{EXP}$ Ref.\cite{gsiprl10}    &$T_{1/2}^{FSUGold}$     &$T_{1/2}^{NL3}$      &$T_{1/2}^{TM1}$ \\
\noalign{\smallskip}\hline\noalign{\smallskip}

$^{289}114$   &	    9.87 (3)	& 10.06 (3)   &	  $ 0.97^{+0.97}_{-0.32}~s$ 	            &$	0.35^{+0.08}_{-0.06}~s	$&$	0.33^{+0.07}_{-0.06}~s	$&$0.31^{+0.07}_{-0.05}~s	$\\

$^{285}112$   &	    9.21 (3)	& 9.39 (3)   &	  $ 30^{+30}_{-10}~s$ 	            &$	6.40^{+1.48}_{-1.20}~s	$&$	6.05^{+1.41}_{-1.13}~s	$&
$5.65^{+1.31}_{-1.05}~s	$\\

$^{281}Ds$   &	    8.73 (3)	& 8.90 (3)   &	  $ 140^{+510}_{-90}~s$ 	            &$	43.61^{+10.67}_{-8.60}~s	$&$	39.18^{+9.66}_{-7.67}~s	$&
$38.08^{+9.39}_{-7.45}~s $\\

$^{288}114$   &	    9.95 (3)	& 10.14 (3)   &	  $ 0.47^{+0.24}_{-0.12}~s$ 	            &$	0.09^{+0.01}_{-0.02}~s	$&$	0.08^{+0.02}_{-0.01}~s	$&
$0.08^{+0.01}_{-0.02}~s $\\
\end{tabular}
\end{ruledtabular}
\end{table*}

    The research work of P. Roy Chowdhury is sponsored by the UGC (No.F.4-2/2006(BSR)/13-224/2008(BSR)) under Dr. D.S. Kothari Postdoctoral Fellowship Scheme. The author (P.R.C.) is thankful to Prof. M.M. Sharma for his support in this work. The works of A. Bhattacharyya and G. Gangopadhyay are partially supported by UGC (UPE $\&$ DRS), CSIR and UGC (DRS) respectively.

\end{document}